\documentclass[11pt,twoside]{article}
\usepackage{./asp2014}

\aspSuppressVolSlug
\resetcounters

\begin{document}

\title{A study of the Lyman-$\alpha$ line profile in DBA white dwarfs}
\author{C. Genest-Beaulieu and P. Bergeron
\affil{Universit\'e de Montr\'eal, Montr\'eal, Qu\'ebec, Canada, \email{genest@astro.umontreal.ca},\email{bergeron@astro.umontreal.ca}}}

\paperauthor{C. Genest-Beaulieu}{genest@astro.umontreal.ca}{}{Universit\'e de Montr\'al}{D\'epartement de Physique}{Montr\'eal}{Qu\'ebec}{H3C 3J7}{Canada}
\paperauthor{P. Bergeron}{bergeron@astro.umontreal.ca}{}{Universit\'e de Montr\'eal}{D\'epartement de Physique}{Montr\'eal}{Qu\'ebec}{H3C 3J7}{Canada}

\begin{abstract}
The hydrogen abundances in DBA white dwarfs determined from optical or
UV spectra have been reported to differ significantly in some
studies. We revisit this problem using our own model atmospheres and
synthetic spectra, and present a theoretical investigation of the
Lyman-$\alpha$ line profile as a function of effective temperature and
hydrogen abundance. We identify one possible solution to this
discrepancy and show considerable improvement from a detailed analysis
of optical and UV spectra of DBA stars.
\end{abstract}

\section{Introduction}

The most widely used technique to determine the atmospheric parameters
of DB white dwarfs is the so-called spectroscopic technique
\citep[see, e.g.,][]{Bergeron2011}. By comparing the line shapes of
the normalized observed spectrum of a star with the predictions from
model atmospheres, it is possible to obtain the effective temperature,
$T_{\rm eff}$, and the surface gravity, $\log g$, of a particular
star. An example of the spectroscopic technique is shown in the left
panel of Figure \ref{Fig1}. The hydrogen abundance $N({\rm H})/N({\rm
  He})$ in a DBA star can also be constrained if a spectrum of the
H$\alpha$ line is available (see insert in the left panel of
Fig.~\ref{Fig1}).

\articlefigure[width=0.91\textwidth]{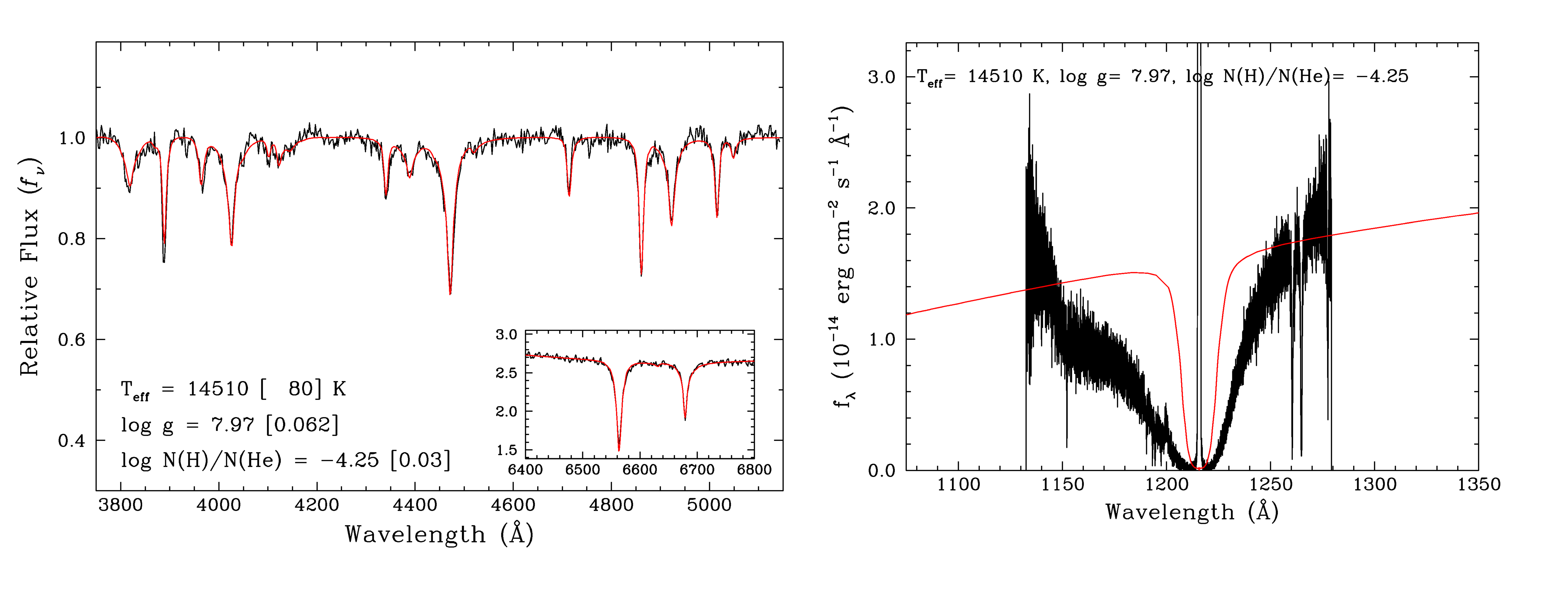}{Fig1}{\emph{Left:}
  Best fit to the optical spectrum of G200-39 (WD
  1425+540). \emph{Right:} Predicted Lyman-$\alpha$ line using the
  atmospheric parameters determined from optical spectroscopy (red
  line), compared to the observed Lyman-$\alpha$ line for G200-39
  (black line).}

On the right panel of Figure \ref{Fig1}, we plotted the Lyman-$\alpha$
profile predicted from the synthetic spectrum obtained from chemically
homogeneous models with the atmospheric parameters determined from
optical spectroscopy together with the observed Lyman-$\alpha$ line of
G200-39 (HST spectrum kindly provided to us by Siyi Xu). As is
clearly shown, the predicted absorption feature is substantially
weaker than the actual observed line. The use of the spectroscopic
technique on this part on the spectrum would thus yield very different
atmospheric parameters than what is obtained from the H$\alpha$
line. Although we are only showing here the discrepancy between the
optical and UV spectra for G200-39, this problem is also present
in other DBA white dwarfs we studied (usually hotter), but to a lesser extent.

\section{Exploring the parameter space}

The strength and shape of the Lyman-$\alpha$ line are affected by both
the effective temperature and the hydrogen abundance. To get a
stronger absorption feature, we can either lower $T_{\rm eff}$ or
increase $N({\rm H})/N({\rm He})$. To know by how much we would
need to lower the effective temperature, we fitted the observed
Lyman-$\alpha$ line by forcing the surface gravity and hydrogen
abundance to the optical values ($\log g=7.97$ and $\log
N({\rm H})/N({\rm He})=-4.25$), allowing only $T_{\rm eff}$ to vary. Our
best fit is shown in the left panel of Figure \ref{Fig2}. The right
panel of Figure \ref{Fig2} shows the same experiment, but by forcing
the surface gravity and effective temperature to the optical values
($T_{\rm eff,opt}=14,510$ K) and allowing only the hydrogen abundance
to vary.

\articlefigure[width=0.92\textwidth]{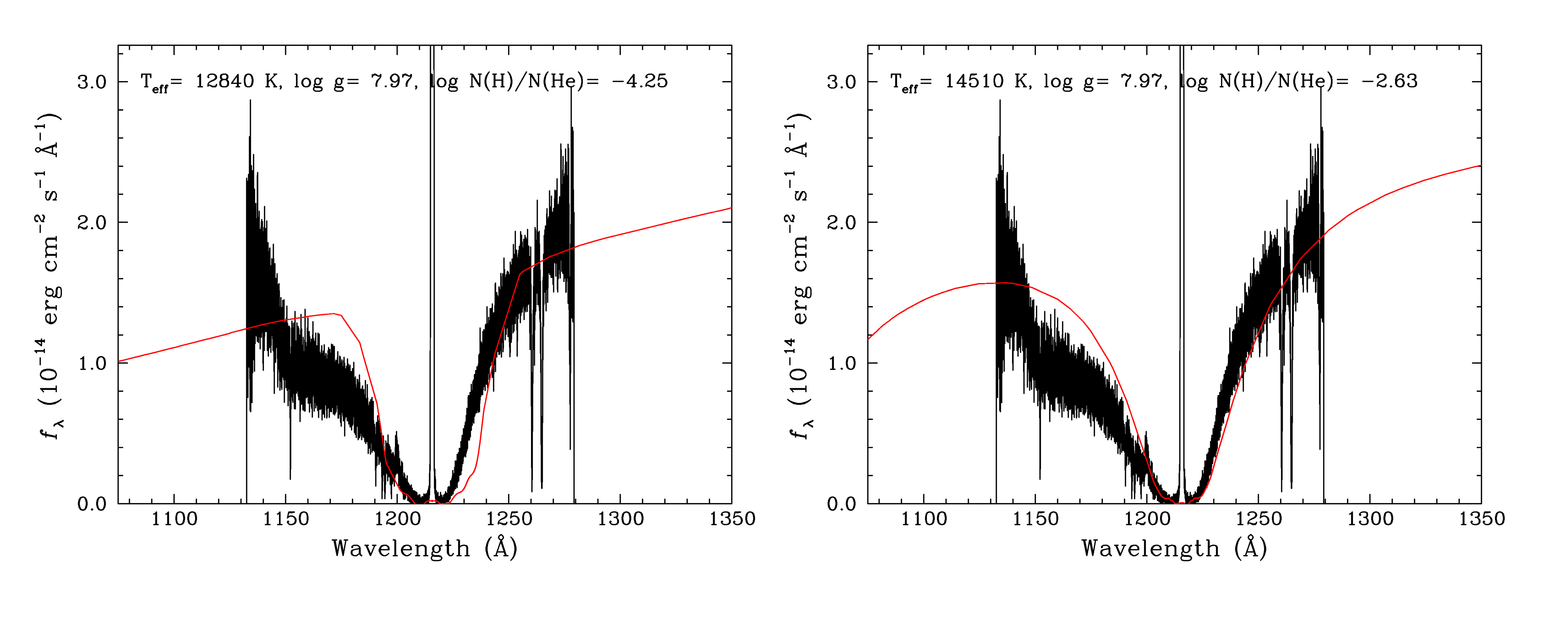}{Fig2}{\emph{Left:}
  Best fit to the Lyman-$\alpha$ line by allowing the effective
  temperature to vary, but by forcing the values of the surface
  gravity and hydrogen abundance obtained from the optical
  fit. \emph{Right:} Same as left panel, only the hydrogen abundance
  is allowed to vary and the effective temperature if forced to the
  optical value.}

In order to get a strong enough Lyman-$\alpha$ line for G200-39,
we would need to lower the effective temperature by about 1600 K,
which is in strong disagreement with the optical temperature ($T_{\rm
  eff,opt}=14,510$ K vs $T_{\rm eff,UV}=12,840$ K); this
temperature difference is clearly larger than the measurement
errors. We get approximately the same effect on the strength of the
Lyman-$\alpha$ line by increasing the hydrogen abundance by 1.62 dex
($\log N({\rm H})/N({\rm He})=-4.25$ from the optical fit vs $\log
N({\rm H})/N({\rm He})=-2.63$ from the UV fit), again in strong
disagreement with the optical value. Note, however, 
that the {\it shape} of the line profile predicted with a higher
hydrogen abundance is in better agreement with the observations than
the prediction with a lower effective temperature.

\section{An inhomogeneous H/He abundance distribution?}

The Lyman-$\alpha$ line is produced higher in the atmosphere than its
optical counterparts. An inhomogeneous atmosphere, where the hydrogen
abundance increases as we approach the star's surface, could explain
the discrepancy between the ultraviolet and the optical spectroscopic
fits. To test the effect of a chemically inhomogeneous atmosphere on
the predicted synthetic spectrum, we fixed the hydrogen abundance in
the convective zone and then obtained the H/He abundance as a function
of depth from the diffusive equilibrium approximation
\citep{MacDonald1991}. An example of such a profile for the
atmospheric parameters of G200-39 is shown in the left panel of Figure
\ref{Fig3}. Note that in this exploratory calculation, we used a
homogeneous atmospheric structure (temperature and pressure), with
parameters obtained from the optical solution (left panel of
Fig.~\ref{Fig1}), and the inhomogeneous profile was only used when
calculating the synthetic spectrum. The resulting Lyman-$\alpha$ line
for G200-39, displayed in the right panel of Figure \ref{Fig3}, is in
much better agreement with the observations.

\articlefigure[trim=0 7cm 0
  0,width=0.92\textwidth]{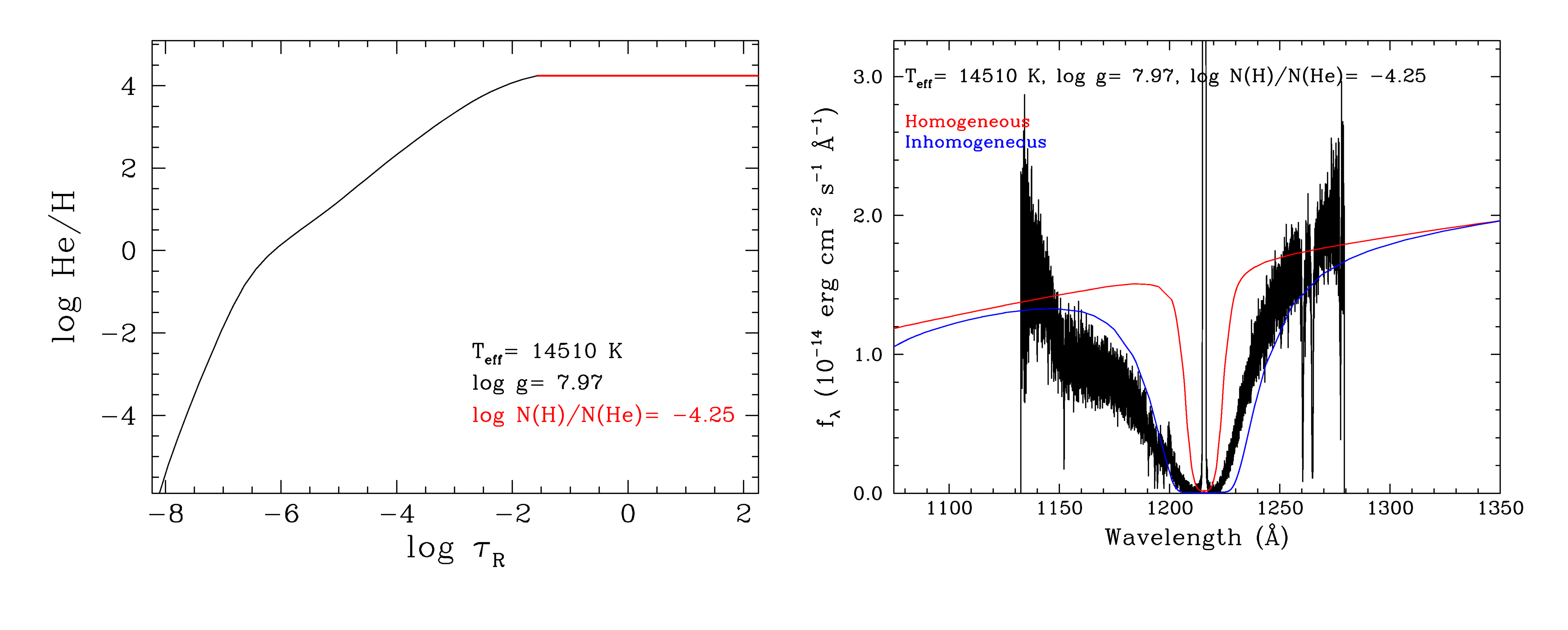}{Fig3}{\emph{Left:}
  Hydrogen abundance profile as a function of Rosseland optical depth
  in an inhomogeneous atmosphere at $T_{\rm eff} = 14,510$ K and $\log
  g = 7.97$. The hydrogen abundance in the convection zone
  (represented by the thick red line) is fixed at $\log
  N({\rm H})/N({\rm He}) = -4.25$. \emph{Right:} Lyman-$\alpha$ line
  predicted from chemically homogeneous (red) and inhomogeneous (blue)
  synthetic spectra, compared to the observed spectrum of G200-39.}

\articlefigure[angle=270,trim=3cm 4cm 3cm
  4cm,width=0.66\textwidth]{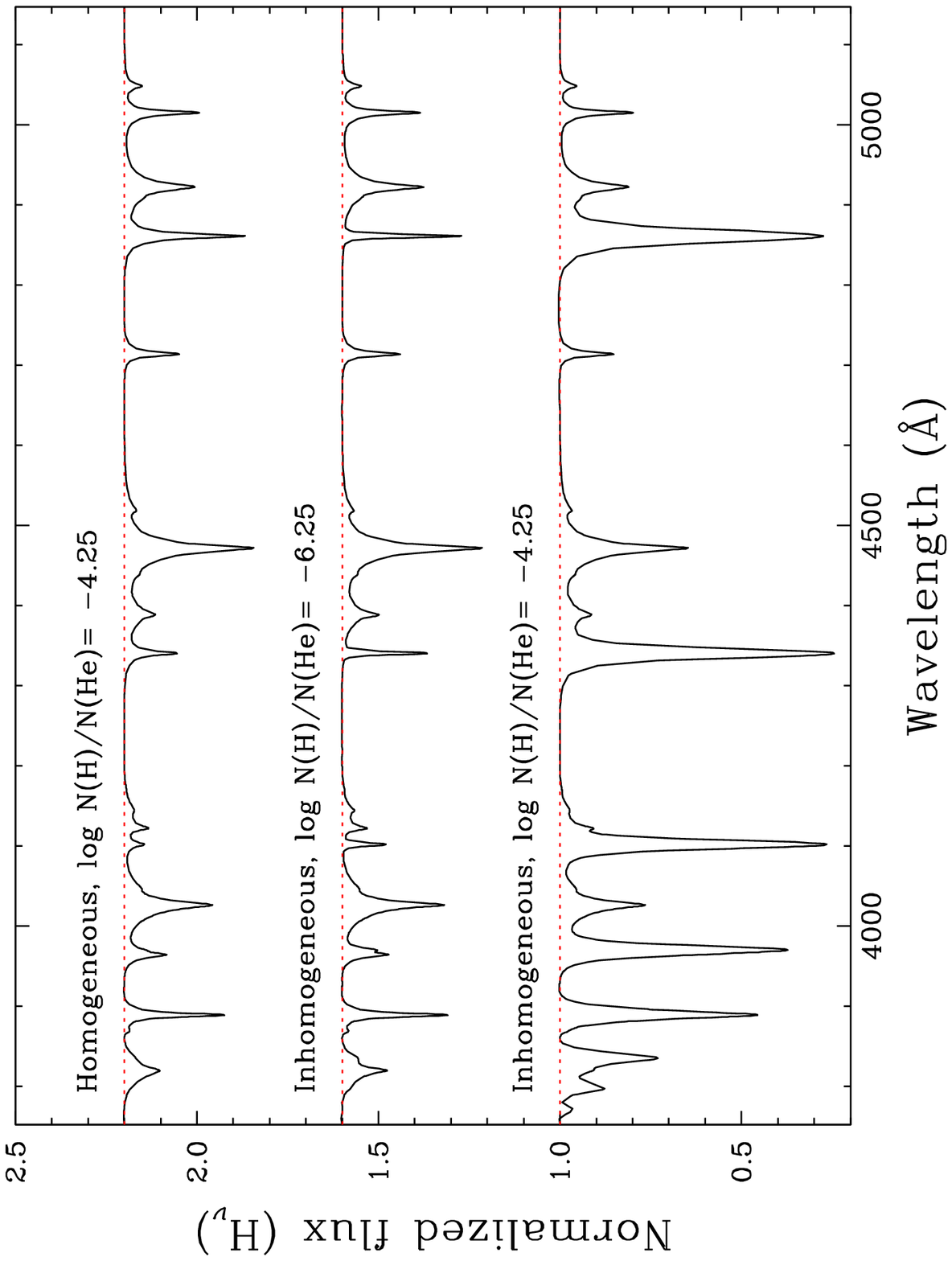}{Fig4}{\emph{Top:}
  Synthetic (normalized) spectrum ($T_{\rm eff} = 14,510$ K, $\log g = 7.97$, and
  $\log N({\rm H})/N({\rm He}) = -4.25$), obtained by using a homogeneous H/He
  abundance profile. \emph{Center:} Synthetic spectrum ($T_{\rm eff} =
  14,510$ K, $\log g = 7.97$, and $\log N({\rm H})/N({\rm He}) = -6.25$), obtained
  by an inhomogeneous H/He abundance profile. \emph{Bottom:} Same as
  top spectrum, but obtained from an inhomogeneous H/He abundance
  profile.}

The effect of using an inhomogeneous abundance profile on the optical
spectrum is illustrated in Figure \ref{Fig4} for a model
at $T_{\rm eff} = 14,510$ K, $\log g = 7.97$, and $\log
N({\rm H})/N({\rm He}) = -4.25$ in the convection zone (bottom
spectrum). This spectrum can be compared with that obtained from a
chemically homogeneous atmosphere with the same hydrogen abundance
(top spectrum). As can be seen, the hydrogen lines in the
inhomogeneous model are predicted much stronger than in the
homogeneous model, which we recall, is representative of the
observed spectrum of G200-39. In order to produce hydrogen lines of
comparable strength, the hydrogen abundance in the convection zone of
the inhomogeneous model needs to be reduced by a factor of 100 (middle
spectrum). These results offer the tantalizing possibility that the
hydrogen abundances in DBA white dwarfs determined from homogeneous
models have been largely overestimated, and consequently that the
total hydrogen mass present in these stars has been significantly
overestimated as well.

\section{Conclusions}

The hydrogen abundance determined from spectroscopic fits to H$\alpha$
is much smaller than that obtained from the Lyman-$\alpha$ line, when
using homogeneous model atmospheres. To correctly fit the
Lyman-$\alpha$ line, we either need to lower the temperature, or to
use a larger $N({\rm H})/N({\rm He})$ ratio than the value obtained from
fits to the optical spectrum. Our exploratory calculations suggest
that this discrepancy can be resolved by using inhomogeneous H/He
abundance profiles, with hydrogen abundances in the convection zone
significantly smaller than those inferred from homogeneous models. If
these results are confirmed, the total hydrogen mass present in DBA
stars may be grossly overestimated.

Our next step is to implement this inhomogeneous hydrogen abundance
profile in the calculation of the atmospheric structure itself (and
not only in the synthetic spectrum) to get a self-consistent solution.

\acknowledgements 
We are deeply grateful to Siyi Xu for letting us use her HST spectrum
of G200-39. This work was supported in part by the NSERC Canada and by the Fund FRQ-NT
(Qu\'{e}bec).

\end{document}